\documentclass[preprint,prl,floatfix]{revtex4}
\usepackage{times,graphicx,subfigure}
\graphicspath{{./}}

\def\img{\mathbf i}
\def\ee{\mathbf e}

\def\rv{{\bf r}}
\def\qv{{\bf q}}

\def\qs{q^{*}}
\def\q0{q_{0}^{*}}
\def\Rg0{R_{g0}}
\def\chie{\chi_{e}}
\def\barchia{\chi^{*}_{a}}

\begin{document}

% title page
\title{Fluctuations in Symmetric Diblock Copolymers:
       Testing A Recent Theory}
\author{Jian Qin and David C. Morse}

\affiliation{
Department of Chemical Engineering and Materials Science, 
University of Minnesota, 
421 Washington Ave. S.E., Minneapolis, MN 55455}

\date{\today}

\begin{abstract}
Composition fluctuations in disordered melts of symmetric diblock 
copolymers are studied by Monte Carlo simulation over a range of 
chain lengths and interaction strengths. Results are used to test 
three theories: (1) the random phase approximation (RPA), (2) the 
Fredrickson-Helfand (FH) theory, which was designed to describe 
large fluctuations near an order-disorder transition (ODT), and 
(3) a more recent renormalized one-loop (ROL) theory, which 
reduces to FH theory near the ODT, but which is found to be 
accurate over a much wider range of parameters.
\end{abstract}

\maketitle

%-------------------------------------------------------------------------
% Introduction
%-------------------------------------------------------------------------

Diblock copolymers are linear polymers that contain two blocks of 
chemically distinct monomers. Diblock copolymer melts can exhibit 
both a disordered phase and a variety of periodic ordered phases.
\cite{BatesFredrickson1999} Composition fluctuations in the 
disordered phase can be measured by small angle x-ray and neutron 
scattering experiments.  The structure factor $S(q)$ measured in such 
experiments is well approximated for very long polymers far from any 
transition by the random phase approximation \cite{Leibler_1980}
(RPA). The RPA for $S(q)$ is based on the same physical approximations 
as those underlying the self-consistent field (SCF) theory that is 
often used to describe the ordered phases.
Several closely related coarse-grained theories that attempt to 
improve upon the RPA 
\cite{Fredrickson_Helfand_1987, Barrat_Fredrickson_1991,
Holyst_Vilgis_93, Wang_2002, Beckrich_Wittmer_2007,
Wittmer_Beckrich_2007, Morse_2006, Piotr_Morse_2007, Qin_Morse_2009, 
Morse_Qin_2011, Qin_Morse_2011} yield corrections that decrease 
with increasing chain length $N$, and thus reduce to the RPA in
the limit $N \rightarrow \infty$. The simulations presented 
here were designed to test the quantitative accuracy of these 
theories.

%-------------------------------------------------------------------------
%Theories
%-------------------------------------------------------------------------

We consider the structure factor $S(q)$ in a dense liquid of symmetric 
AB diblock copolymers, each of $N$ monomers. Let
\begin{equation}
   S(q) \equiv \int d\rv \; 
   \ee^{\img \qv\cdot\rv}
   \langle \delta \psi(\rv) \delta \psi(0) \rangle ,
\end{equation}
where $\delta \psi(\rv) \equiv \delta c_{A}(\rv) - \delta c_{B}(\rv)$ 
and $q \equiv |\qv|$. $\delta c_{i}(\rv)$ is the deviation of the 
number concentration $c_{i}(\rv)$ of $i$ monomers from its spatial 
average. In a diblock copolymer melt, $S(q)$ is maximized at a nonzero 
wavenumber $\qs$.

The RPA predicts \cite{Leibler_1980} an inverse correlation function
of the form
\begin{equation}
  cN S_0^{-1}(q) = F(q \Rg0) - 2 \chie N
  ,
  \label{eq:RPA}
\end{equation}
where $\Rg0 = (N/6)^{1/2}b$ is the radius of gyration of a random walk 
polymer with statistical segment length $b$, and $c$ is the total 
monomer concentration. The parameter $\chie$ is an effective interaction 
parameter that is used in the RPA and SCF theories to characterize the 
degree of incompatibility.  Here, $F(x)$ is a known dimensionless 
function \cite{Leibler_1980} that has a minimum at a value $x^{*}$, 
yielding a maximum in $S_{0}(q)$ at a corresponding wavenumber 
$\q0 \equiv x^{*}/\Rg0$. The subscript of `$0$' is used to denote RPA 
predictions for the quantities $S(q)$, $R_{g}$ and $\qs$. For symmetric 
diblock copolymers, $F(x^{*}) = 21.99$ and $x^{*} = 1.95$, so that $S_0(q)$ 
diverges at a predicted spinodal value $(\chie N)_{s} = 10.495$. 

A series of closely-related coarse-grained theories have attempted to 
improve upon the RPA by taking into account fluctuation effects that 
it neglects.
\cite{Fredrickson_Helfand_1987, Barrat_Fredrickson_1991,
Holyst_Vilgis_93, Wang_2002, Beckrich_Wittmer_2007,
Wittmer_Beckrich_2007, Morse_2006, Piotr_Morse_2007, Qin_Morse_2009, 
Morse_Qin_2011, Qin_Morse_2011} Here, we compare simulation results to 
the Fredrickson-Helfand \cite{Fredrickson_Helfand_1987} (FH) theory, 
which was the first such theory for diblock copolymers, and the
renormalized one-loop (ROL) theory, 
\cite{Piotr_Morse_2007, Qin_Morse_2009, Morse_Qin_2011, Qin_Morse_2011}
which is the most recent. These theories all yield predictions for 
$S^{-1}(q)$ as a sum $S^{-1}(q) = S^{-1}_{0}(q) + \delta S^{-1}(q)$, 
in which the correction $\delta S^{-1}(q)$ is proportional to a 
small parameter $\bar{N}^{-1/2}$ with $\bar{N} \equiv Nb^6c^2$. 
Physically, $\bar{N}^{1/2} = N^{1/2}b^{3}c$ is a measure of overlap: 
It is proportional to the number $\bar{N}^{1/2} \sim R^{3}/V$ of chains 
of excluded volume $V=N/c$ that can pack in the volume $R^{3}$ pervaded 
by a chain of size $R \sim \sqrt{N}b$.  

The FH theory for $S(q)$ was designed to describe the dominant effects 
of strong composition fluctuations very near the order-disorder 
transition (ODT) of a melt of long, symmetric diblock copolymers. It 
yields a simple nonlinear equation for the peak intensity $S(\qs)$. 
This can also be expressed in terms of an ``apparent" interaction 
parameter $\barchia$ that we define by fitting $S(\qs)$ to 
the RPA, by taking
\begin{equation}
   c N S^{-1} (\qs) \equiv 2 [  (\chie N)_{s} - \barchia N ],
   \label{eq:chiadef}
\end{equation}
where $(\chie N)_{s} = 10.495$. The FH theory predicts 
\begin{equation}
  cN \delta S^{-1}(q) = \bar{N}^{-1/2}
  \frac{B}{ [ (\chie N)_{s} - \barchia ]^{1/2} } ,
  \label{dSinvFH}
\end{equation}
with $B = 280$. 

More recently, several groups have developed a family of renormalized 
one-loop (ROL) theories of correlations in polymer liquids
\cite{Wang_2002, Beckrich_Wittmer_2007, Wittmer_Beckrich_2007, 
Morse_2006, Piotr_Morse_2007, Qin_Morse_2009, Morse_Qin_2011, Qin_Morse_2011} 
with a potentially wider range of validity. The characteristic
feature of these theories is the development of methods to 
distinguish non-universal effects of monomer-scale correlations
from the universal effects of long-wavelength correlations, and 
to absorb the effects of short-wavelength correlations into
renormalized values of the RPA parameters $b$ and $\chi_{e}$. 
Beckrich, Wittmer and colleagues 
in Strasbourg \cite{Beckrich_Wittmer_2007, Wittmer_Beckrich_2007} 
have developed a theory for universal deviations from random-walk 
statistics in homopolymer melts, and verified their predictions by 
extensive simulations.  Wang \cite{Wang_2002} gave predictions for 
$S(q)$ near $q=0$ in homopolymer blends. Our group 
\cite{Piotr_Morse_2007, Morse_Chung_2009, Qin_Morse_2009, Morse_Qin_2011, Qin_Morse_2011} 
has developed predictions for $S(q)$ and single-chain correlations 
in both blends and diblock copolymer melts for arbitrary $q$, which 
reduce to the results of other authors in appropriate limits. The 
ROL theory for symmetric diblock copolymers
\cite{Piotr_Morse_2007, Morse_Qin_2011, Qin_Morse_2011} 
predicts a correction of the form
\begin{equation}
   cN\delta S^{-1}(q) = \bar{N}^{-1/2} H(q\Rg0, \barchia N) 
   \quad, \label{dSinvROL}
\end{equation}
in which the dimensionless function $H$ is defined a sum of Fourier 
integrals that we evaluate numerically. 
This expression has been shown \cite{Morse_Qin_2011, Qin_Morse_2011} 
to approach the FH prediction of Eq.  (\ref{dSinvFH}) very near the 
ODT.  This ROL prediction also appears
\cite{Piotr_Morse_2007, Morse_Qin_2011} to be the first correction 
to the RPA within a systematic expansion of $c N\delta S^{-1}(q)$ in 
powers of $\bar{N}^{-1/2}$. Because the validity of this expansion is 
not restricted to the vicinity of the ODT, we expect the ROL theory 
(but not FH theory) to remain accurate for $\bar{N} \gg 1$ even far 
from the ODT.  

The limitations of the RPA have been well documented by previous 
simulations of diblock copolymer melts 
\cite{Binder_Fried_1991a, %Binder_Fried_1991b, 
Larson_1996, Grest_Kremer_1999, Matsen_Vassiliev_2006} 
and experiments.
\cite{Bates_Rosedale_1990, Almdal_Bates_1992, Rosedale_Bates_1995}
Both experiments and simulations show a decrease in $\qs$ with
decreasing temperature $T$ and a nonlinear dependence of the peak 
intensity $S(\qs)$ on $1/T$ near the ODT that are not predicted 
by the RPA.
Previous simulations have not, however, provided very precise tests 
of the absolute accuracy of the RPA or (particularly) of theories 
that predict corrections to the RPA. One reason for this has been the 
absence of a clear prescription for relating the interaction 
parameter $\chie$ that is required as an input to these coarse-grained 
theories to the more microscopic parameters that are controlled in a 
simulation. The analysis presented here uses several methods to avoid 
or minimize this ambiguity.

%-------------------------------------------------------------------------
% Model
%-------------------------------------------------------------------------
Our simulations use a potential energy similar to that of Grest 
and coworkers. \cite{Grest_Kremer_1999} Non-bonded beads interact 
via a purely repulsive Lennard-Jones potential, with 
$u_{ij}(r) = 
\varepsilon_{ij} \left[ 4(r/\sigma)^{-12} - 4(r/\sigma)^{-6} + 1 \right]$
for $r < r_{c}$, with $r_{c} = 2^{1/6}\sigma$.
Consecutive beads within each chain interact via a harmonic bond potential 
$\kappa (r - l_{0})^{2}/2$.  All simulations discussed here use parameters 
$\varepsilon_{AA} = \varepsilon_{BB} = kT$, 
$l_{0} = \sigma$, and $\kappa = 400 kT \sigma^{-2}$, with a total 
concentration $c = 0.7 \sigma^{-3}$. The magnitude $\epsilon_{AB}$ of the 
$AB$ repulsion is controlled by a parameter 
$\alpha = \varepsilon_{AB} - \varepsilon_{AA}$, which indirectly controls 
$\chie$.

We have simulated melts of chains of $N=$16, 32, 64, and 128 beads 
in a periodic $L \times L \times L$ cubic box. Two values of $L$
were used for each chain length $N$, to monitor finite size effects. 
For each $N$ and $L$, we conducted parallel replica-exchange MC 
simulations \cite{Earl_Deem_2005} of systems with different values 
of $\alpha$ at constant $T$. To sample configurations, we combined 
a hybrid MD/MC move,\cite{Mehlig_1992} in which short constant energy 
MD simulations are used to generate proposed MC moves, with reptation 
and double-rebridging \cite{Banaszak_dePablo_2003} configuration bias 
moves. 

To test the RPA, FH, and ROL theories, one needs an unambiguous way to 
determine values for the parameters $b$ and $\chie$ that these theories 
require as inputs. Our comparisons with theory all use a value of $b$ 
defined, as by the Strasbourg group,\cite{Wittmer_Beckrich_2007} as a 
limit $b^{2} \equiv \lim_{N \rightarrow \infty} 6R_{g}^{2}(N)/N$.  This 
has been evaluated by numerically extrapolating results for $R_{g}^{2}(N)$ 
for homopolymer melts ($\alpha = 0$) of varying $N$, giving 
$b = 1.41 \sigma$. This definition of $b$ is required for consistency 
with the ROL theory, which predicts that random-walk statistics 
\cite{Beckrich_Wittmer_2007} and the RPA for $S(q)$ 
\cite{Piotr_Morse_2007} with renormalized values of $b$ and $\chie$ 
are exact only in the limit $N \rightarrow \infty$.  An analogous 
procedure for estimating for $\chie(\alpha)$, discussed below, is also 
used in some comparisons.

%-----------------------------------------------------------------------
% Correlation at $\alpha = 0$
%-----------------------------------------------------------------------
\textit{Limit $\chie = 0$:}
We first consider the special case of a system with $\alpha = 0$, 
or $\epsilon_{AB} = \epsilon_{AA}$, corresponding to $\chie = 0$.
This is an idealization of a neutron scattering experiment in which 
A and B blocks are labelled by differential deuteration, but are 
otherwise identical.  
In this limit, both the RPA and ROL predictions for $S(q)$ depend
only on intramolecular correlation functions for chains in a dense
melt. The RPA, which assumes random walk statistics, predicts 
$cNS^{-1}_{0}(q) = F(q \Rg0)$. The ROL theory yields predictions 
that differ from the RPA only as a result of predicted 
${\cal O}(\bar{N}^{-1/2})$ deviations from random walk statistics.
\cite{Qin_Morse_2011}

\begin{figure}[tb] \centering
  \includegraphics[width=0.60\textwidth,height=!]{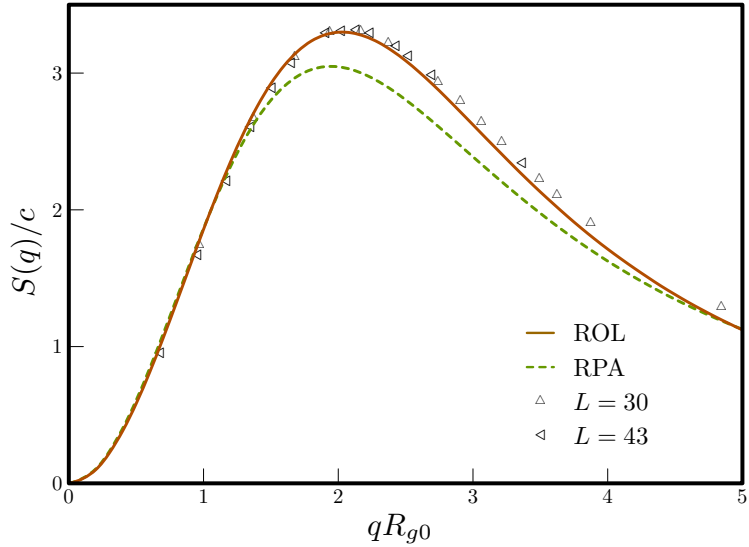}
  \caption
  {$q$ dependence of $S(q)$ for $N=64$, at $\alpha = 0$, for 
  $L = 30$ and $L = 43$. The RPA and ROL predictions are 
  shown by dashed and solid lines, respectively.}
  \label{fig:N64alpha0}
\end{figure}

Fig. \ref{fig:N64alpha0} shows simulation results and predictions
for $S(q)$ vs. $q \Rg0$ in systems with $N=64$ $(\bar{N} = 240$) 
and $\alpha = 0$. Here, $\Rg0^{2} \equiv Nb^{2}/6$, with $b =
1.41\sigma$. Equivalent results are obtained for box sizes $L = 30$ 
and $L = 43$. The RPA (dashed line) slightly underestimates both 
the peak amplitude $S(\qs)$ and peak wavenumber $\qs$. 
The ROL prediction (solid line), however, fits the data almost 
perfectly, with no adjustable parameters. 
We find a similar level of agreement at $\alpha = 0$ for other 
chain lengths.
  
\begin{figure}[tb] \centering
  \includegraphics[width=0.60\textwidth,height=!]{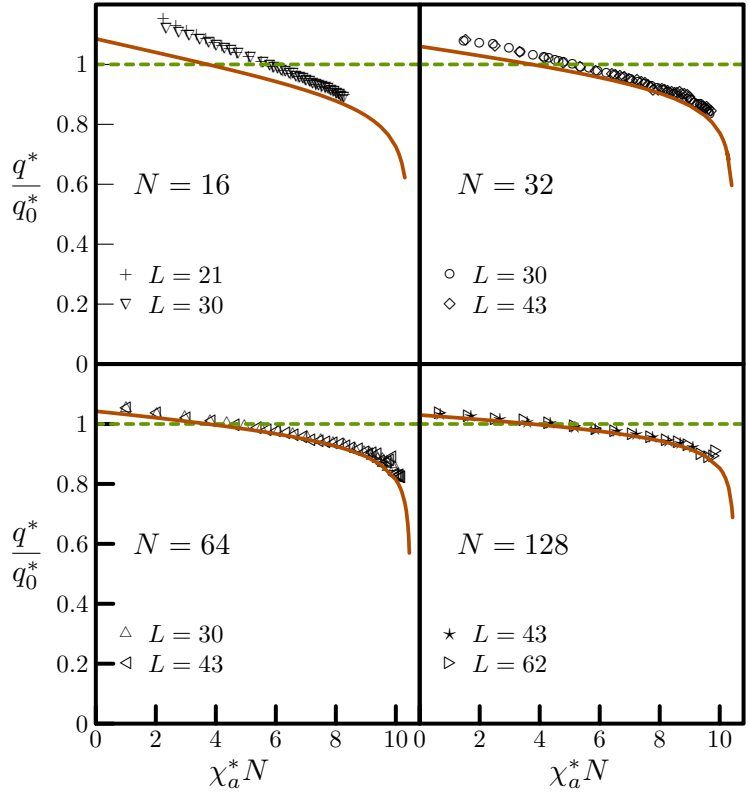}
  \caption{Peak wavenumber $\qs$, normalized by the RPA prediction
   $\q0$, {\it vs} $\barchia N$ for $N=$16, 32, 64, and 128.  Data for 
   two values of box size $L$ are shown for each chain length $N$. ROL 
   predictions are solid lines.}
  \label{fig:qstarchi}
\end{figure}

%-----------------------------------------------------------------------------
% Peak position
%-----------------------------------------------------------------------------
\textit{Peak wavenumber:}
We next discuss the evolution of peak wavenumber $\qs$ with 
changes in $\alpha$, or $\chie(\alpha)$. Fig. \ref{fig:qstarchi}
shows our results for $\qs/ \q0$, where $\q0 = 1.95/\Rg0$ is 
the RPA prediction. The ratio $\qs/\q0$ is plotted vs. the 
quantity $\barchia N$ that is defined in Eq. (\ref{eq:chiadef}) 
by fitting the observed peak intensity to the RPA. This plot 
thus shows peak wavenumber plotted vs. a measure of peak 
intensity. 

\begin{figure}[tb]
  \begin{center}
  \subfigure{(a)\label{fig:SinvN64chi}}{\includegraphics[width=0.60\textwidth,height=!]{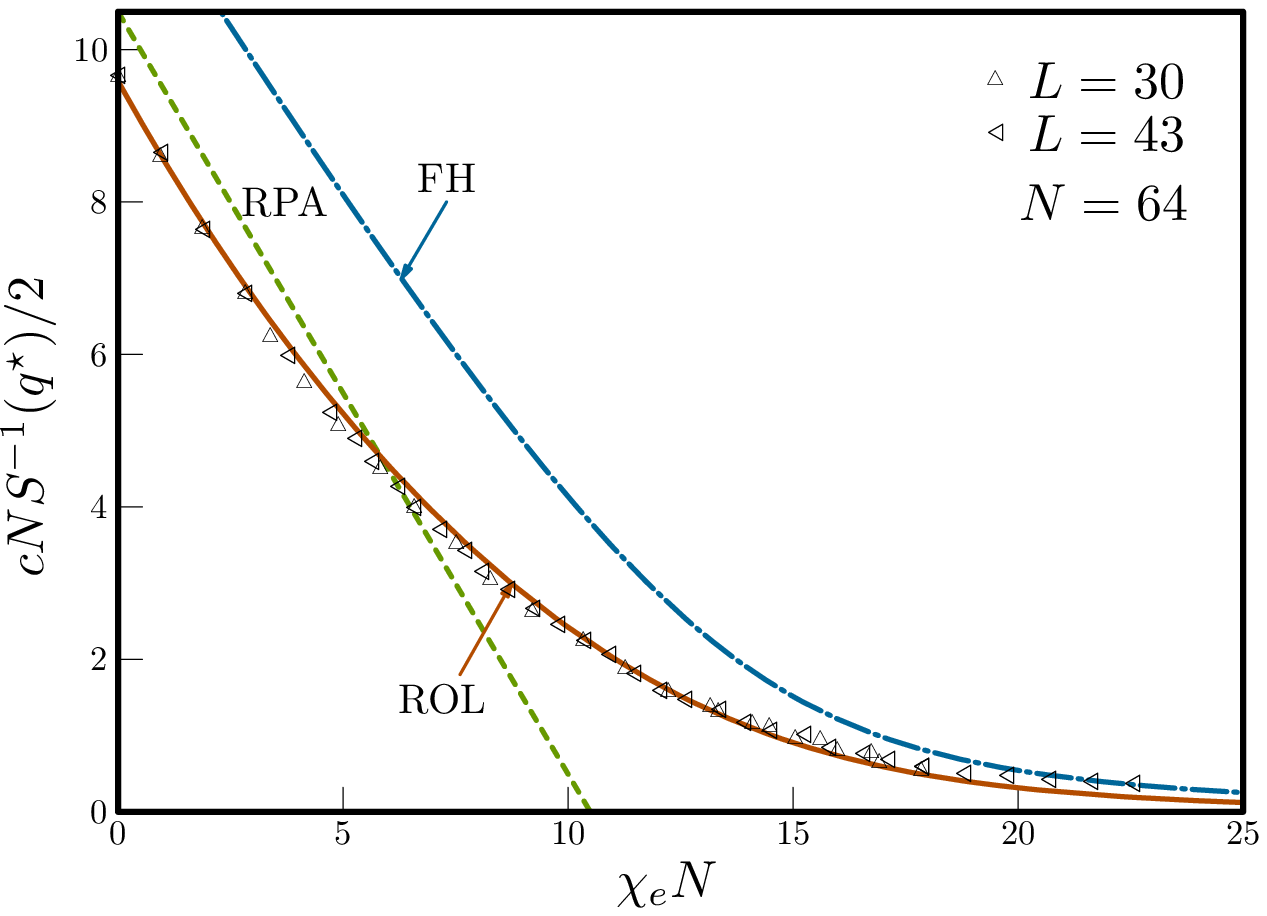}}
  \subfigure{(b)\label{fig:SinverseN}}{\includegraphics[width=0.60\textwidth,height=!]{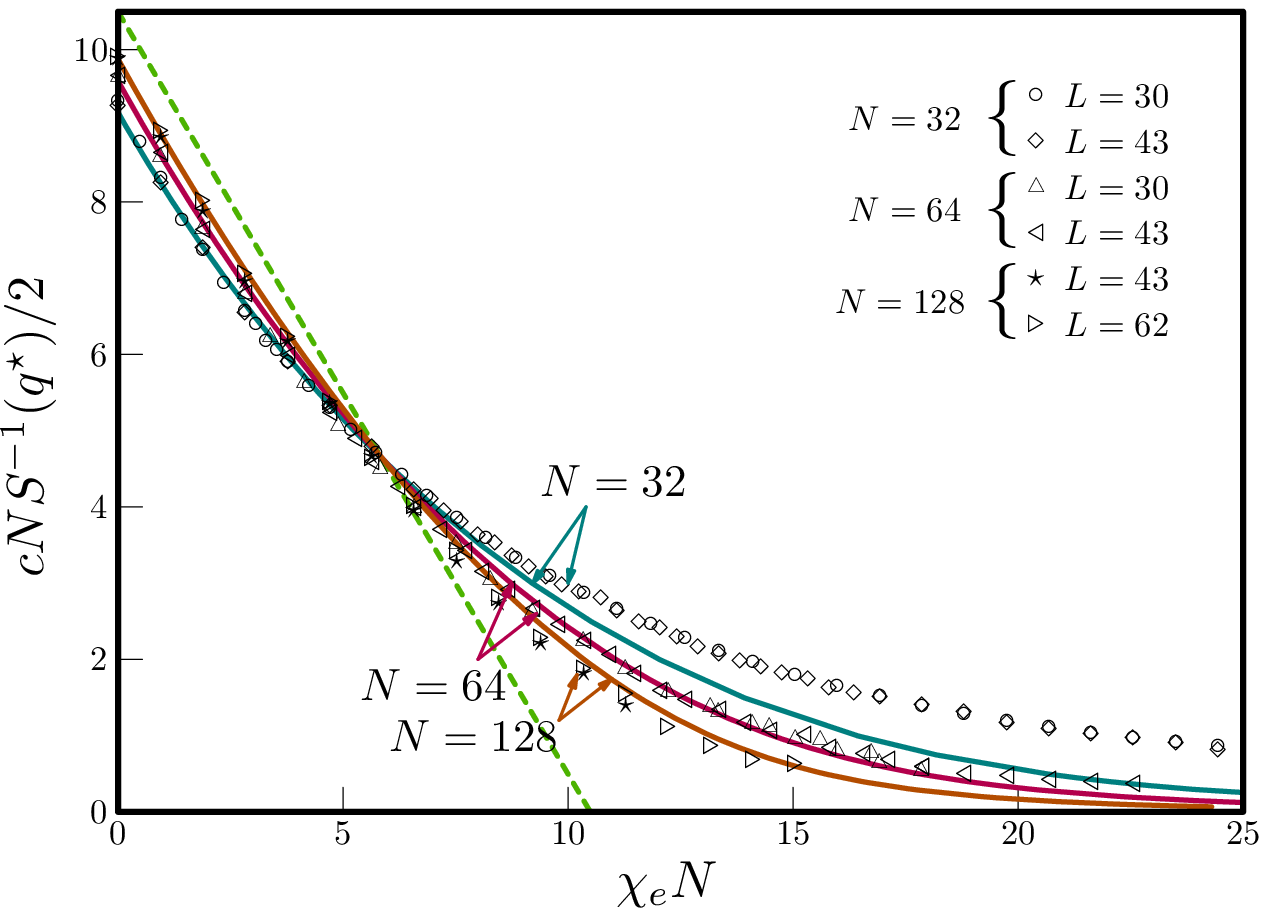}}
  \caption
  {Inverse peak intensity {\it vs.} $\chie N$ for $N = 64$ (panel a, upper)
   and for $N=$ 32, 64, and 128 (panel b, lower). For each $N$, results are
   show for two values of the box size $L$. $\chie$ is estimated using Eq.
   (\ref{chiealpha}), using $z_c = 0.2965$. Straight dashed lines show the 
   RPA prediction. Solid lines are ROL predictions. The dot-dashed curve 
   in panel a is the FH prediction. }
   \label{fig:Sinverse}
   \end{center}
\end{figure}

Both $\qs$ and $S(\qs)$ were determined by fitting values of 
$S(q)$ at a discrete set of allowed wavevectors to a smooth 
function. For $N=$64 and 128 and large values of $\alpha$, this 
procedure becomes unreliable because there are too few allowed 
values of $q$ within the peak. In Figs. \ref{fig:qstarchi} and 
\ref{fig:Sinverse}, we thus show results for $\qs$ and $S(\qs)$ 
for each $N$ and $L$ only over the range of values of $\alpha$ 
for which the peak remains broad enough to allow a reliable 
fit. Within these regions, results from different box sizes 
are consistent, and we see no evidence of a transition to an 
ordered phase.

Simulation results for $\qs$ decrease monotonically with 
increasing $\barchia$ or $\alpha$, exhibiting a total drop of 
15-20 \% over the range shown, in agreement with previous 
results.\cite{Matsen_Vassiliev_2006} 
Note that $\qs/\q0 > 1$ for small $\barchia N$, consistent 
with the behavior shown in Fig. 1 for $\alpha = 0$, and that 
$\qs/\q0 < 1$ only for $\barchia N \gtrsim 5$. 
The RPA predicts a constant $\qs/\q0 = 1$ (dashed line). 
ROL predictions (solid lines) were obtained from parametric 
plots of predictions for $\qs$ and $S(\qs)$ over a range 
of values of $\chie N$. The ROL theory captures all of 
the observed qualitative features, and is strikingly accurate for 
the longest chains ($N=$ 64 and 128). The predictions appear to
become systematically more accurate with increasing $N$, consistent
with the claim \cite{Piotr_Morse_2007, Qin_Morse_2009} that the 
ROL theory is the first correction to the RPA within an expansion 
in powers of $\bar{N}^{-1/2}$. 

%----------------------------------------------------------------------------
% Mapping $\alpha$ to $\chie$
%----------------------------------------------------------------------------
\textit{Estimating $\chie(\alpha)$:}
In all of the theories considered here, the degree of incompatibility
between A and B monomers is characterized by an effective interaction
parameter $\chie$. To test the predictions of these theories for how 
the peak intensity $S(\qs)$ evolves with increasing $\chie$, we need 
an independent estimate of how $\chie(\alpha)$ depends on $\alpha$. 
(The comparisons shown in Figs. 1 and 2 did not require this.)

We estimate $\chie(\alpha)$ using a method based on thermodynamic 
perturbation theory. Ref. \onlinecite{Morse_Chung_2009} discusses a 
perturbation theory for a structurally symmetric blend of $A$ and $B$ 
homopolymers, both of length $N$, with a pair potential 
$u_{ij}(r) = \epsilon_{ij}u(r)$, with $\epsilon_{AA} = 
\epsilon_{BB}$ and $\alpha = \epsilon_{AB}-\epsilon_{BB}$, in which the 
excess free energy per monomer $f_{ex}(\alpha, N)$ is expanded in powers 
of $\alpha$.  In such a model, $f_{ex}(\alpha = 0, N)=0$.  It was shown 
\cite{Morse_Chung_2009} that $f_{ex}$ is given to ${\cal O}(\alpha)$ by 
$f_{ex}(\alpha, N) \simeq \alpha z(N)\phi_{A}\phi_{B}$, where $\phi_{i}$ 
is the fraction of $i$ monomers, and where
\begin{equation}
  z(N) = \int d\rv \; g(\rv; N) u(\rv) .
\end{equation}
Here, $g(\rv;N)$ is the inter-molecular radial distribution function 
in the reference state with $\alpha = 0$. By comparing this 
perturbation theory to ROL predictions for $S(q)$ at $q = 0$, which 
reduce to the RPA in the limit $N \rightarrow \infty$, while also 
expanding $\chie(\alpha)$ in powers of $\alpha$, it was found that 
the RPA parameter $\chie$ is given to ${\cal O}(\alpha)$ by
\begin{equation}
   \chie \simeq z(\infty)\alpha/kT 
   ,
   \label{chiealpha}
\end{equation}
where $z(\infty) \equiv \lim_{N \rightarrow \infty} z(N)$.  
For the model considered here,\cite{Morse_Chung_2009} we find 
$z(\infty) = 0.2965$. Perturbation theory provides a useful 
description of the homogeneous state of both blends and diblock 
copolymer melts because the critical values of $\alpha$ and 
$\chie$ are proportional to $1/N$, implying that the accuracy 
of perturbation theory must improve with increasing $N$. 

%---------------------------------------------------
% Correlation at $\alpha \neq 0$
%---------------------------------------------------
\textit{Peak intensity:}
Fig. \ref{fig:Sinverse} shows how the inverse peak intensity 
$c NS^{-1}(\qs)$ varies with $\chie N$, plotted using 
Eq. (\ref{chiealpha}) for $\chie(\alpha)$.  Fig. \ref{fig:SinvN64chi} 
compares results for chains with $N=64$ to the RPA 
(straight dashed line), FH (dot-dashed), and ROL (solid) predictions. 
Fig. \ref{fig:SinverseN} compares results for chains of length 
$N=$ 32, 64, and 128 to RPA and ROL predictions. The comparisons 
to theory involve no adjustable parameters.

Several qualitative features are immediately apparent: 
The quantity $cNS^{-1}(\qs)$ is greater than the RPA prediction 
(suppressed fluctuations) for $\chie N \agt 6$, as predicted by the 
FH theory, but is less than the RPA prediction (enhanced fluctuations) 
for $\chie N \alt 6$. For $\chie N \alt 10$, results for different 
$N$ converge towards the RPA prediction with increasing $N$, as 
predicted by both FH and ROL theories.  
In Fig. \ref{fig:SinvN64chi}, the FH and ROL predictions for $N=64$ 
are very similar to each other and to the simulation results in the 
strong fluctuation regime, where $\chie N \agt 15$ and 
$cN S^{-1}(\qs) \alt 1$. This is the regime that the FH theory was 
designed to describe. For $\chie N \alt 15$, however, the FH theory 
fails, while the ROL theory remains remarkably accurate, for this 
chain length, down to $\chie N = 0$.  The ROL theory correctly 
predicts a change in the sign of the deviation of $cN S^{-1}(\qs)$ 
from the RPA prediction at a value of $\chie N \simeq 6$, which is
shown in Fig. \ref{fig:SinverseN} to be almost independent 
of $N$. The FH theory incorrectly predicts a rather large positive 
deviation for all $\chie N \geq 0$. In \ref{fig:SinverseN}, the ROL 
theory is seen to be quite accurate for all three values of $N$ for 
$\barchia N \alt 7$, and to remain accurate for larger values
$\chie N$ for $N=$64 and 128. ROL predictions differ significantly 
from the results only for the shortest chains shown, with $N=$32, 
at large values of $\chie N$. Results for $N=16$ are not shown, 
but exhibit the same trends, and differ even more from the ROL 
prediction. The apparent tendency of the ROL theory to become 
more accurate with increasing $N$ is consistent with the claim 
that it is part of a systematic expansion in powers of 
$\bar{N}^{-1/2}$. 

The comparison shown in Fig. 3 is subject to at least two types of 
error: (i) Errors in the ROL theory itself, due to truncation of 
the loop expansion at first order in $\bar{N}^{-1/2}$, and 
(ii) Errors in our estimate of $\chie$, due to the truncation of 
the Taylor expansion of $\chie(\alpha)$ at first order in $\alpha$. 
The comparisons shown in Figs. 1 and 2 are subject only to type (i). 
The error in $\chie(\alpha)$ is expected to depend only on $\alpha$, 
and to increase with $\alpha$. It should be most serious for large 
values of $\chie N$ and small $N$, because the value of $\alpha$ 
required to obtain a specified value of $\chie N$ increases with 
decreasing $N$.
The fact that the largest discrepancies between the ROL theory and 
the simulation results occurs in Fig. 3 for the shortest chains at
large values of $\chie N$ suggests that our use of a simple linear 
approximation for $\chie(\alpha)$ may be a significant source of 
error for the shortest chains.

%-----------------------------------------------------------------------
% Summary
%-----------------------------------------------------------------------

In summary, we have presented an unusually systematic simulation study of 
how $S(q)$ depends upon both chain length and degree of AB repulsion in AB 
diblock copolymer melts, and compared our results to several theories. 
The use of theoretically motivated procedures for obtaining independent
estimates of the RPA parameters $b$ and $\chie$, defined by extrapolation 
to $N=\infty$, allowed much more precise comparisons to theory than has 
previously been possible. 
The range of values of $\bar{N}$ studied here, $\bar{N} \leq 480$, overlaps 
the lower end of the range studied in experiments on polymers, for which 
$\bar{N}$ is more often $10^{3} - 10^{4}$. 
A recent theory, the ROL theory, is shown to be a substantial improvement 
over the FH theory, with a much wider range of validity. The ROL theory
is strikingly accurate for the longest chain lengths studied, and for 
sufficiently small values of $\chie N$.  All our results appear to be 
consistent with the claim that this theory is the first correction to 
the RPA within a systematic expansion, and should thus become more 
accurate with increasing $\bar{N}$.  The results indicate that theories 
developed to describe subtle deviations from the random-walk and RPA 
theories in liquids of very long polymers, based on an expansion in powers 
of $\bar{N}^{-1/2}$, can accurately describe liquids of surprisingly short 
chains.

\begin{acknowledgments}
This work was supported by NSF grant DMR-097338.
\end{acknowledgments}

% Bibliography
\bibliography{dbcprl}

\end{document}